\documentclass[12pt,reqno]{amsart}
\usepackage{epsfig}
\usepackage{graphicx}
\begin{document}
\baselineskip=0.7cm 
\theoremstyle{plain}
\newtheorem{thm}{Theorem}[section]
\newtheorem{lem}{Lemma}[section]
\newtheorem{prop}{Proposition}[section]
\newtheorem{coll}{Conclusion}
\theoremstyle{remark}
\newtheorem{rem}{Remark}[section]
\title{Folded solitary waves and Foldons in 2+1 dimensions}
\author{Xiao-yan Tang$^1$ and Sen-yue Lou$^{1,2,3}$}
\dedicatory{{$^1$Department of Physics, Shanghai Jiao Tong
University, Shanghai, 200030, P.R.China}\\
{$^2$School of Mathematics, The University of New South Wales,
Sydney, NSW 2052, Australia}\\
$^3$ Department of Physics, Ningbo University, Ningbo 315211, P.
R. China}
 \begin{abstract}
A general type of localized excitations, folded solitary waves and
foldons, are defined and studied both analytically and
graphically. The folded solitary waves and foldons may be
``folded" in quite complicated ways and possess quite rich
structures and abundant interaction properties. The folded
phenomenon is quite universal in the real natural world. The
folded solitary waves and foldons are obtained from a quite
universal formula and the universal formula is valid for some
quite universal (2+1)-dimensional physical models. The
``universal" formula is also extended to a more general form with
many more independent arbitrary functions.

\vskip.1in

\leftline{\bf \emph{PACS.}05.45.Yv, 02.30.Jr, 02.30.Ik.}
\end{abstract}

\maketitle

In the study of nonlinear science, soliton theory plays a very
important role and has been applied in almost all the natural
sciences especially in all the physical branches such as the fluid
physics, condense matter physics, biophysics, plasma physics,
nonlinear optics, quantum field theory and particle physics etc..
Almost all the previous studies of soliton theory especially in
high dimensions are restricted in single valued situations.
However, the real natural phenomena are very complicated. In
various cases, it is even impossible to describe the natural
phenomena by single valued functions. For instance, in the real
natural world, there exist very complicated folded phenomena such
as the folded protein \cite{Protein}, folded brain and skin
surfaces and many many other kinds of folded biologic
systems\cite{skin}. The simplest multi-valued (folded) waves may
be the bubbles on (or under) a fluid surface. Various kinds of
ocean waves are really folded waves also.

To study the complicated folded natural phenomena is very
difficult. Similar to the single valued cases, the first important
question we should ask is: Are there any stable multi-valued
(folded) localized excitations? For convenience later, we define
the multi-valued localized excitations folded solitary waves.
Furthermore, if the interactions among the folded solitary waves
are completely elastic, we call them foldons. In (1+1)-dimensional
case, the simplest foldons are so-called loop solitons\cite{loop}
which can be found in many (1+1)-dimensional integrable
systems\cite{loop} and have been applied in some possible physical
fields like the string interaction with external
field\cite{string}, quantum field theory\cite{field} and particle
physics\cite{particle}. However, in our knowledge, there is no
study at all for the possible higher dimensional foldons. That is
the main topic of this letter.

In (1+1)-dimensions, it has been proven that when a physical model
can be expressed by a partial differential equation (PDE), then
under some suitable approximations, one can always find nonlinear
Sch\"ordinger type equations\cite{Calogero} that is why the
(1+1)-dimensional soliton theory can be successively applied in
almost all the physical branches. Similarly, if (2+1)-dimensional
physical model can be expressed by a PDE, then under some suitable
approximations, one can find the Davey-Stewartson (DS) type
equations\cite{Maccari}.

Recently, it is found that a quite ``universal" formula \cite{vsa}
\begin{eqnarray}\label{0}
U\equiv \frac{-2\Delta q_{y}p_{x}}{(a_0+a_1 p+a_2 q+a_3 p q)^2},\
\qquad \Delta\equiv a_0a_3-a_1a_2,
\end{eqnarray}
is valid for suitable fields or potentials of various
(2+1)-dimensional physically interesting integrable models
including the DS equation, the dispersive long wave equation
(DLWE)\cite{vsaDLWE}, the Broer-Kaup-Kupershmidt (BKK), and so on.
In Eq. \eqref{0}, $p\equiv p(x,\ t)$ is an arbitrary function of
$\{x,\ t\}$, $q\equiv q(y,\ t)$ may be either an arbitrary
function for some kinds of models such as the DS system or an
arbitrary solution of a Riccati equation for some others, say, the
DLWE while $a_0,\ a_1,\ a_2$ and $a_3$ are taken as constants. One
of the most important results obtained from \eqref{0} is that for
all the models mentioned above there are quite rich localized
excitations such as the solitoffs, dromions, lumps, breathers,
instantons, ring solitons, peakons, compactons, localized chaotic
and fractal patterns and so on\cite{vsa}.

Now the natural and important question is: Can we find some types
of (2+1)-dimensional foldons from the ``universal" formula?
Fortunately, the answer is obviously positive because of the
arbitrariness of the functions $p$ and $q$ included in the
universal formula. To find more general types of foldons, we
extend the universal formula to a more general form for the DLWE
before to give out concrete foldons. For the (2+1)-dimensional
DLWE
\begin{equation}\label{dlwe}
u_{yt}+\eta_{xx}+u_x u_y+u u_{xy}=0,\ \eta_t+u_x+\eta u_x+u
\eta_x+u_{xxy}=0,
\end{equation}
\eqref{0} is valid for the field $v\equiv -\eta-1$ and the related
exact solution reads
\begin{equation}\label{dlwv}
v\equiv -\eta-1=U,\  u=\pm\frac{2p_x(a_1+a_3q)}
{a_0+a_1p+a_2q+a_3pq}+u_0,
\end{equation}
with $p$ being an arbitrary functions of $\{x,\ t\}$, $q=q(y,t)$
being an arbitrary solution of the Riccati equation,
$q_t-a_0c_0-(a_1c_1+a_2c_0-a_0c_2)q-(a_3c_1-a_2c_2)q^2=0$, and
$u_0 = -p_x^{-1}[p_t\pm
p_{xx}-a_0c_1-(a_1c_1+a_2c_0+a_0c_2)p-(a_1c_2+a_3c_0)p^2].$ The
(1+1)-dimensional DLWE ($y=x$ of \eqref{dlwe}) is also called the
classical Boussinesq equation. There exist a large number of
papers to discuss the possible applications and exact solutions of
the (1+1)-dimensional DLWE\cite{1+1}. Various interesting
properties of the (2+1)-dimensional DLWE have been studied by many
authors\cite{dlwe}--\cite{dlwenoP}.


To extend the universal formula to a more general form, we use the
multi-linear variable separation approach again. For the DLWE
\eqref{dlwe}, the transformation $ \{u=u_0\pm 2\frac{f_x} {f},\
\eta=2\frac{f_{xy}}{f} -2\frac{f_xf_y}{f^2}-1\}$ with $u_0\equiv
u_0(x,t)$ degenerates two equations of \eqref{dlwe} to a trilinear
form
\begin{eqnarray}\label{tri}
& &[f_{xxxy}\pm (f_{xyt}+u_{0x} f_{xy}+u_0 f_{xxy})]f^2-[(f_{xxx}
\pm u_0 f_{xx}\pm f_{xt}) f_y \pm
f_x f_y u_{0x} \nonumber\\
& &+(f_{xxy}\pm 2 u_0 f_{xy}\pm f_{yt}) f_x+f_{xy} f_{xx}\pm f_t
f_{xy}]f+2 f_x f_y(f_{xx}\pm u_0 f_x \pm f_t)=0.
\end{eqnarray}
To solve the trilinear equation \eqref{tri}, one has to use some
prior ansatz. The variable separation solution \eqref{dlwv} is
resulted from the ansatz $f=a_0+a_1p+a_2q+a_3pq$.

For the DLWE \eqref{dlwe}, there are two sets of infinitely many
symmetries and every symmetry possesses an arbitrary function of
$t$ or $y$\cite{dlwesym}. That means infinitely many arbitrary
functions of $y$ and $t$ can be entered into the solutions of
\eqref{dlwe}. So it is possible to extend the solution
\eqref{dlwv} to a more general one with more arbitrary functions.
After finishing the detailed calculations, we find that the
following variable separation ansatz,
\begin{eqnarray}\label{ansatz}
f=q_0+\sum_{i=1}^Np_iq_i,
\end{eqnarray}
where $\{q_i,\ i=0,\ 1,\ 2,\ ...,\ N\}$, and $\{p_i,\ i=1,\ 2,\
...,\ N\}$ are functions of $\{y,\ t\}$ and $\{x,\ t\}$
respectively, solves the trilinear equation \eqref{tri} under the
conditions
\begin{eqnarray}
q_{it}=\sum_{j=0}^N(c_{i,j}+q_iC_j)q_j,\
p_{it}=(c_{00}-u_0\partial_x-\partial_x^2)p_i-c_{0i}+\sum_{j=1}^N(c_{j0}p_i-c_{ji})p_j,
\label{pit}
\end{eqnarray}
for all $i=0,\ 1,\ 2,\ ...,\ N$ and $\{c_{ij},\ i,j=0,\ 1,\ 2,\
...,\ N,\ C_j,\ j=1,\ 2,\ ...,\ N\}$ being arbitrary functions of
$t$. Obviously, the general ansatz \eqref{ansatz} will return back
to the known one for $N=2,\ q_0=a_0, \ q_1=a_1,\ q_2=a_2q,\
q_3=a_3q,\ p_1=p_3=p, \ p_2=1$.

The corresponding solution for the field $v=-\eta-1$ now reads
\begin{equation}\label{dlwV}
v=\frac{-2\sum_{i=1}^Np_{ix}q_{iy}}{q_0+\sum_{i=1}^Np_iq_i}
+\frac{2\sum_{i=1}^Np_{ix}q_{i}\left(q_{0y}+\sum_{j=1}^Np_jq_{jy}\right)}
{(q_0+\sum_{i=1}^Np_iq_i)^2}\equiv U_E
\end{equation}
while the quantity $u$ is given by $
u=\pm({2\sum_{i=1}^Np_{ix}q_i})/( {q_0+\sum_{i=1}^N p_iq_i})+u_0.
$ It is clear that in addition to a (1+1)-dimensional arbitrary
function of $\{x,\ t\}$ (one of $u_0$ and $p_i$), $(N+1)(N+2)-1$
arbitrary functions of $t$, $c_{ij},\ C_j$ have been included in
the general solution. Furthermore, various arbitrary functions of
$y$ and $\{x, \ t\}$ will also be included in \eqref{dlwV} after
the coupled systems of \eqref{pit} being solved. Because of the
complexity, we have to leave these problems for our future
studies. Here we just write down the simplest nontrivial case for
the later uses. It is quite trivial that, when we fix $N=1$,
$c_{ij}=C_i=0, \ p_1=p$ and $q_0\rightarrow a_0+q_0$, the formula
\eqref{dlwV} is simplified to
\begin{equation}\label{dlwV1}
v=\frac{2p_{x}\left(q_{1}q_{0y}-(a_0+q_0)q_{1y}\right)}{(a_0+q_0+pq_1)^2}
\end{equation}
with $q_0$ and $q_1$ being arbitrary functions of $y$ and $p$
being an arbitrary function of $\{x,\ t\}$. We can prove that the
simplified quantity expressed by the right hand side of
\eqref{dlwV1} does work for all the known models that allow the
universal formula \eqref{0}. It is interesting also that though we
have not yet proven the validity of \eqref{dlwV} for all the
models mentioned in \cite{vsa}, we do have proven that the
extended form \eqref{dlwV} is really valid at least for some of
them such as the BKK system, the (2+1)-dimensional Burgers system
and the (1+1)-component AKNS system. For instance, the quantities
$g\equiv -2G=U_E$ and $H=(\ln f)_x+u_0/2$ with the same $f,\ p_i$
and $q_i$ determined by \eqref{ansatz} and \eqref{pit} solve the
BKK system $\{H_{ty}-H_{xxy} +2(HH_x)_y +2G_{xx}=0,\
G_t+G_{xx}+2(HG)_{x}=0\}$.

As pointed out in \cite{vsa}, because of the arbitrariness of the
functions appeared in the universal formula and its extended
forms, various class of localized excitations can be constructed.
Now we concentrate on how to find some types of folded solitary
waves and foldons from the universal formula and its extended
forms. First of all, we write a localized function, $px$, in the
form
\begin{eqnarray}
px\equiv \sum_{j=1}^Mf_j (\xi+v_jt),\
x=\xi+\sum_{i=1}^Mg_j(\xi+v_jt),\label{lp1}
\end{eqnarray}
where $v_1<v_2<\cdots v_M$ are all arbitrary constants and
$\{f_j,\ g_j\},\ \forall j$ are all localized functions with the
properties $f_j(\pm \infty)=0,\ g_i(\pm \infty)=
G_i^{\pm}=consts.$ From the second equation of \eqref{lp1}, we
know that $\xi$ may be a multi-valued function in some possible
regions of $x$ by selecting the functions $g_j$ suitably. So the
function $px$ may be a multi-valued function of $x$ in these
regions though it is a single valued function of $\xi$. It is also
clear that $px$ is an interaction solution of $M$ \em localized
\rm excitations because of the property $ \xi|_{x\rightarrow
\infty}\rightarrow \infty$. Actually, most of the known multi-loop
solutions are the special cases of \eqref{lp1}. Now if we take all
the arbitrary functions appeared in the universal formula
\eqref{0} and/or its slightly general one \eqref{dlwV1} (and/or
even its more general extended one \eqref{dlwV}) possess the forms
similar to \eqref{lp1}, then we can get various types of folded
solitary waves and/or foldons.

In Fig. 1, four typical special folded solitary waves are plotted
for the quantity $v$ shown by \eqref{dlwV1} with $p_x=px$,
$p=\int^\xi p_xx_\xi {\rm d\xi}$ and the functions $q_0$ and $q_1$
being given in a similar way
\begin{eqnarray}
q_{\delta y}=\sum_{j=1}^{M_\delta}Q_{\delta j}(\theta),\
y=\theta+R(\theta),\ q_\delta=\int^\theta q_{\delta y} y_\theta
{\rm d \theta},\ (\delta=0,1).\label{lq1}
\end{eqnarray}
In \eqref{lq1}, $Q_{\delta j}(\theta),\ \forall j$ and $R(\theta)$
are localized functions of $\theta$. The more detailed function
selections of the figures are directly given in the figure
caption.

\input epsf
\begin {figure}
\centering \epsfxsize=7cm\epsfysize=5cm\epsfbox{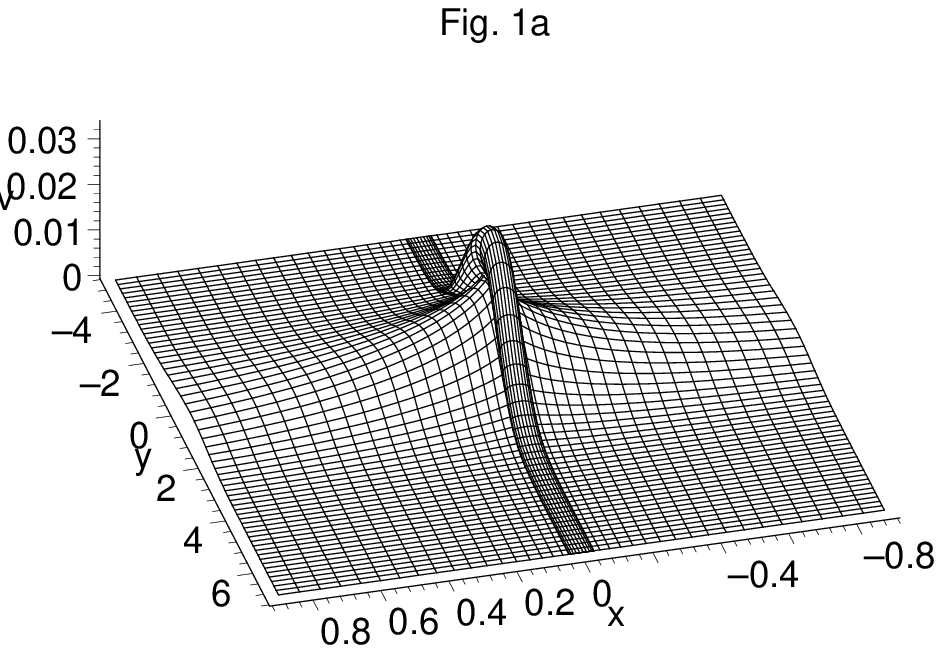}
\epsfxsize=7cm\epsfysize=5cm\epsfbox{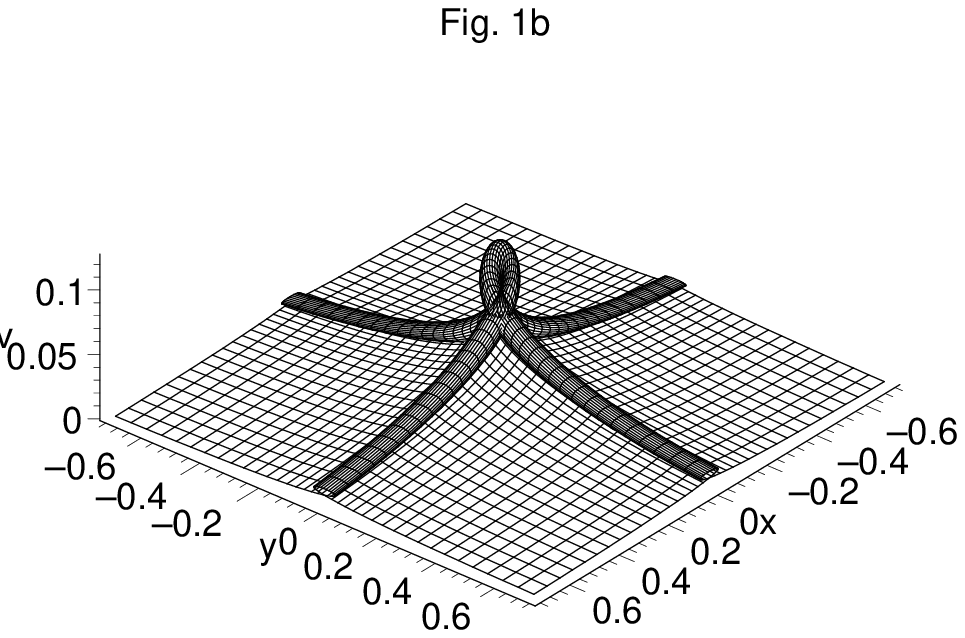}
\epsfxsize=7cm\epsfysize=5cm\epsfbox{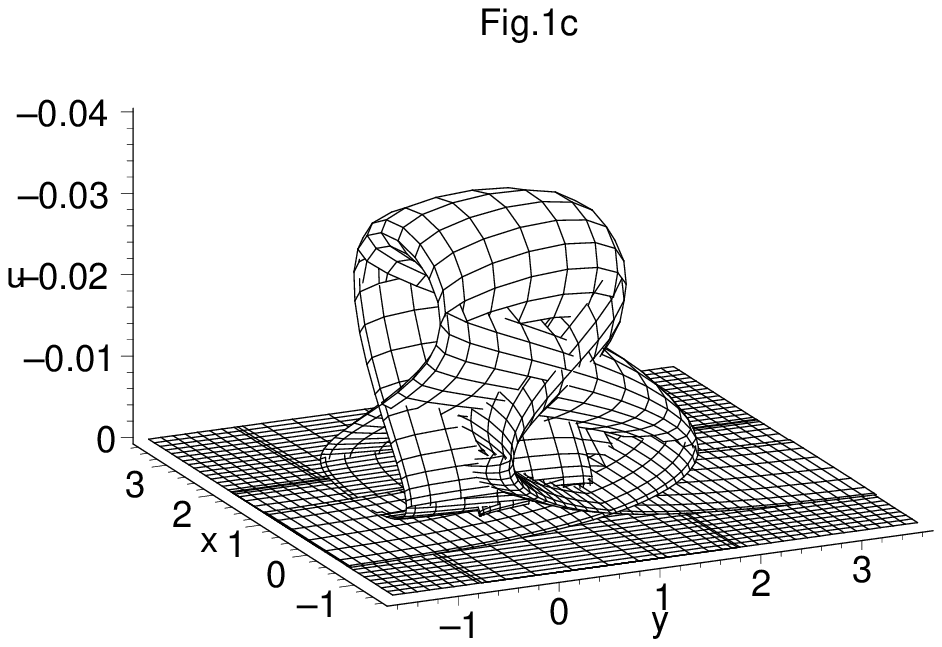}
\epsfxsize=7cm\epsfysize=5cm\epsfbox{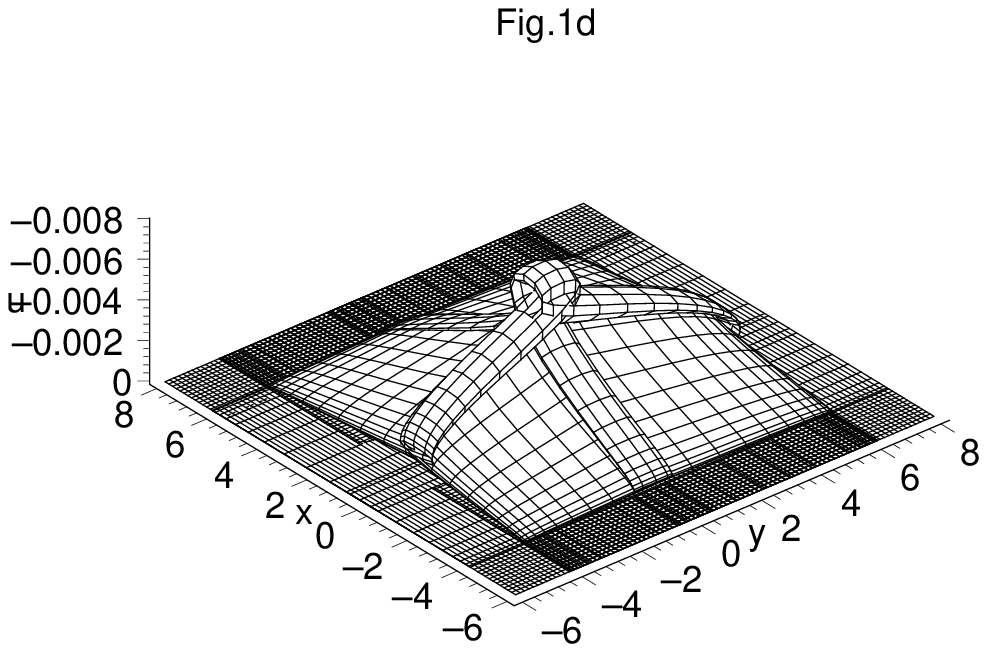} \caption{Four
typical folded solitary wave solutions for the field $v$ expressed
by \eqref{dlwV1} at $t=0$ with \eqref{lp1}, \eqref{lq1} and the
related concrete selections are: (a) $p_x={\rm sech}^2
(\xi-v_1t),\ q_{0y}={\rm sech}^2 \eta,\ q_1=1,\ x=\xi-1.15 \tanh
(\xi-v_1t),\ y=\eta+ l_0\tanh\eta$, $l_0=1$ and $a_0=8$; (b) same
as (a) but with $l_0=-1.15$ and $a_0=4$; (c) $p_x$ and $q_{0y}$
are same as (a), however, $q_{1y}={\rm sech }\eta$,
$x=\xi+2\tanh(\xi-v_1t)+\tanh^2(\xi-v_1t)-k_0\tanh^3(\xi-v_1t)$,
$y=\eta+2\tanh\eta+\tanh^2\eta-k_0\tanh^3\eta$, $k_0=-5.4$ and
$a_0=50$; (d) same as (c) but with $k_0=10$ and $a_0=250$. }
\end{figure}

To construct the possible localized excitations with completely
elastic interaction behaviors, especially, foldons, we should
discuss the asymptotic behaviors of \eqref{dlwV1} and \eqref{lp1}
when $t\rightarrow \pm \infty$. Assuming further that
$F_j(\omega_j)\equiv F_j(\xi+v_jt)\equiv \int f_j{\rm
dx}|_{\omega_j \rightarrow \pm \infty} \rightarrow F_j^{\pm}$, we
can gaze at the $j$th excitation to see the interaction properties
among the localized excitations expressed by \eqref{dlwV1} with
\eqref{lp1} and \eqref{lq1}. In other words, we can take
$\omega_j$ invariant and then take $t\rightarrow \infty$ because
$q_0$ and $q_1$ have been fixed as $t$-independent. The results
read
\begin{eqnarray}\label{limit}
v|_{t\rightarrow \mp \infty} \rightarrow
\frac{2f_j(\omega_j)[q_1q_{0y}-(a_0+q_0)q_y]}{[a_0+q_0+q_1(F_j(\omega_j)+\Omega_j^{\mp})]^2},
\ x|_{t\rightarrow \mp \infty} \rightarrow \xi+ \Gamma_j^{\mp}+
g_j(\xi+v_jt), \label{v-+}
\end{eqnarray}
where $ \Omega_j^\mp=\sum_{i>j}F_i^\mp+\sum_{i<j}F_i^\pm$ and $
\Gamma_j^\mp=\sum_{i>j}G_i^\mp +\sum_{i<j}G_i^\pm$. From the
asymptotic results \eqref{limit}, we know that, (i) the $j$th
localized excitation given by \eqref{dlwV1} with \eqref{lp1} and
\eqref{lq1} is a travelling wave moving in the velocity $v_j$
along the negative ($v_j>0$) or positive ($v_j<0$) $x$-direction;
(ii)  the multi-valued properties (i.e., the structures) of the
$j$th localized excitation is only determined by $g_j$ of
\eqref{lp1}; (iii) the shape of the $j$th excitation will be
changed if $\Omega_j^+\neq \Omega_j^-$, however, it will preserve
its shape if $\Omega_j^+ = \Omega_j^-$; (iv) the total phase shift
for the $j$th excitation is $\Gamma_j^{+}-\Gamma_j^{-}$.

From the above discussions, it is seen that to construct some
(2+1)-dimensional localized soliton type excitations with
completely elastic interaction behaviors becomes an easy task. The
only thing one has to do is to select suitable localized functions
$F_j$ (and then $f_j$) with $\Omega_j^+ = \Omega_j^-$ in
\eqref{lp1}! The localized soliton like solutions can still
possess quite rich structures. Especially, various types of known
single-valued localized excitations are also in the catalog due to
the arbitrariness of the functions included in \eqref{lp1} and
\eqref{lq1}. For instance, if we select $f_i$ and $Q_i$ of
\eqref{lp1} and \eqref{lq1} possesses the peakon shape
solutions\cite{vsa,peak} and/or the compacton shape
solutions\cite{compac} with the property $\Omega_i^+ =
\Omega_i^-$, then we can find some (2+1)-dimensional peakon type
and/or compacton type solutions with completely elastic
interaction properties.

Figs. 2a--2c are plotted to show the possible existence of foldons
which are given by \eqref{dlwV1} with $q_1=1$, \eqref{lp1} and
\eqref{lq1} for $\delta=0$ and the concrete function selections
are also given in the figure caption. From Figs. 2a and 2c, we can
see that the interaction of two foldons is completely elastic.
Because one of the velocities of foldons has been selected as
zero, it can also be seen that there are phase shifts for two
foldons. Especially, before the interaction, the static foldon
(the large one) is located at $x=-1.5$ and after the interaction,
the large foldon is shifted to $x=1.5$.

\input epsf
\begin {figure}
\centering \epsfxsize=7cm\epsfysize=5cm\epsfbox{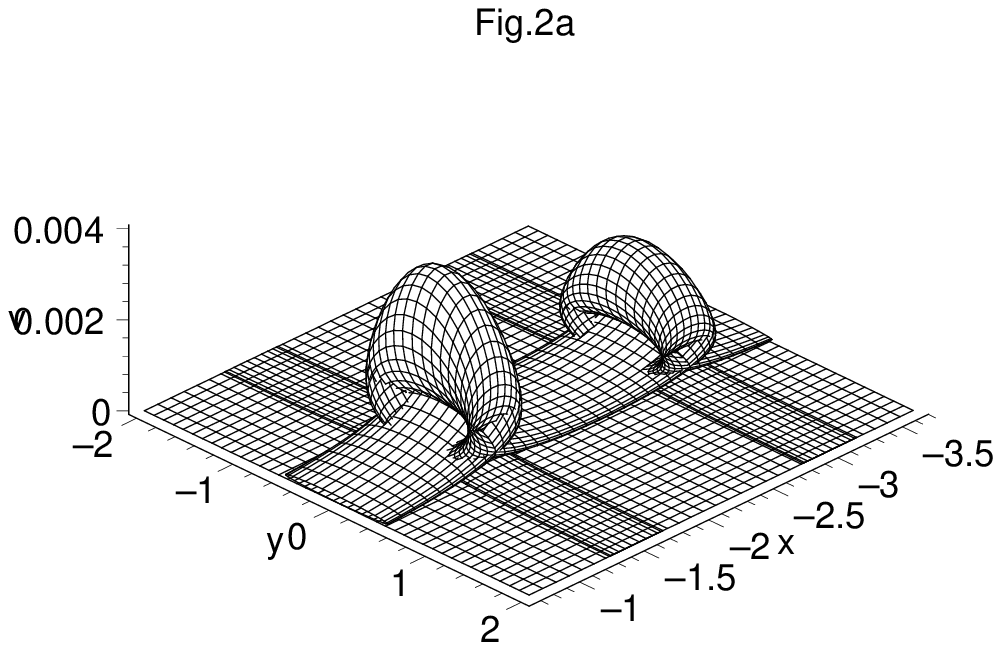}
\epsfxsize=7cm\epsfysize=5cm\epsfbox{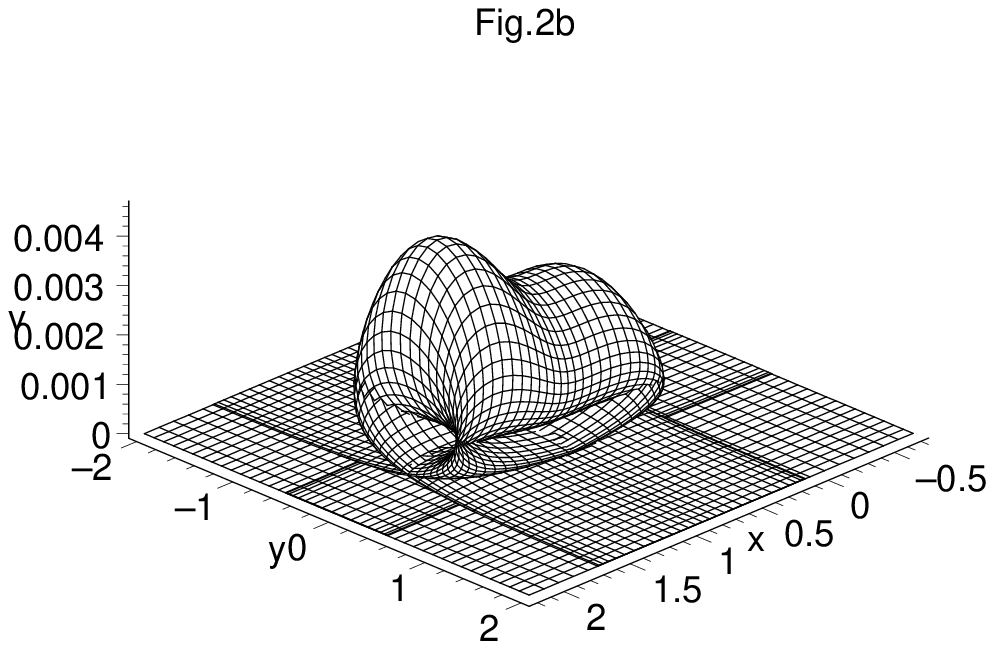}
\epsfxsize=7cm\epsfysize=5cm\epsfbox{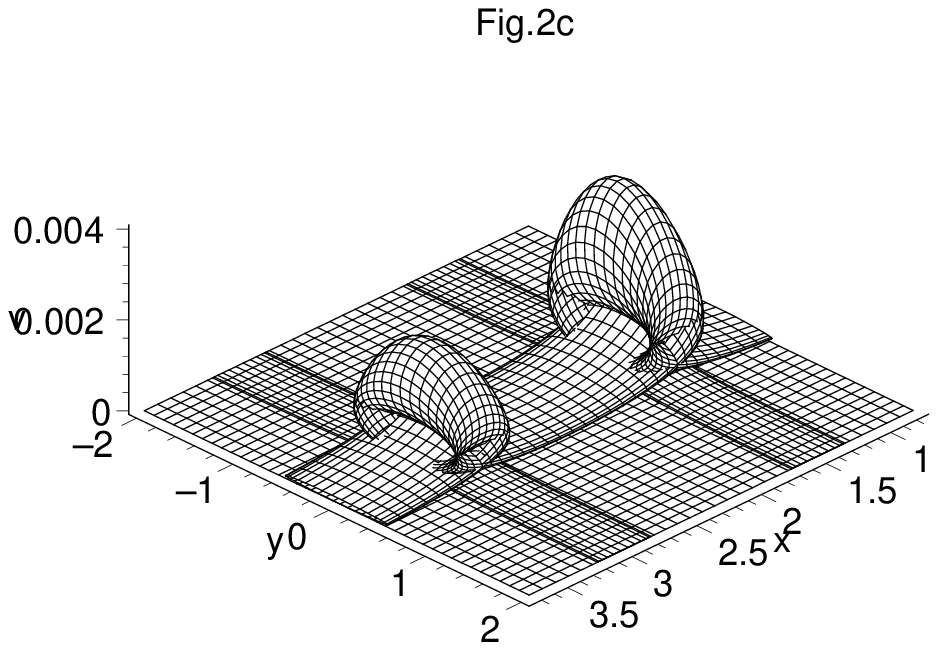}
 \caption{Evolution plots of two foldons for the field $v$ expressed by
\eqref{dlwV1} with $p_x=0.8{\rm sech}^2\xi+0.5{\rm
sech}^2(\xi-0.25t),\ x=\xi-1.5\tanh\xi-1.5\tanh(\xi-0.25t)$,
$q_1=1,\ q_0={\rm sech}^2 \theta$, $y=\theta -2\tanh \theta$ and
$a_0=20$ at times, (a) $t=-18$, (b) $t=7.2$ and (c) $t=18$
respectively.}
\end{figure}

In summary, starting from the ``universal" formula and/or its
extended form, various kinds of localized excitations with and
without completely elastic interaction behaviors can be
constructed easily by selecting the arbitrary functions suitably
according to the asymptotic results \eqref{limit}. Especially, a
new kinds of localized excitations, folded solitary waves and
foldons, are investigated both analytically and graphically. The
foldons may be folded quite freely and complicatedly and then
possess quite rich structures and interaction behaviors. The
explicit phase shifts for all the localized excitations given by
\eqref{dlwV1} with \eqref{lp1} and \eqref{lq1} have been given.

On the one hand, there are a large number of complicated ``folded"
and/or the multi-valued phenomena in the real natural world. On
the other hand, there is no good analytical way to treat these
kinds of complicated phenomena. This letter is only a beginning
attempt, to find some types of possible stable multi-valued
localized excitations, folded solitary waves and foldons, for some
real physical models including some quite ``universal" systems
such as the well known DS equation and the DLWE.

Because the formula \eqref{dlwV1} is valid for various
(2+1)-dimensional interesting models, which are divergently
applied in many physical fields, we believe that the foldons may
be useful in the future studies on the complicated ``folded"
natural world. The more about both the (extended) ``universal"
formula and the general (or special) foldons especially its real
possible applications should be studied further.

The authors are in debt to thank the helpful discussions with
Prof. C. Rogers, Dr. W. Schief, Prof. C-z Qu and Dr. H-c Hu. The
work was supported by the National Outstanding Youth Foundation of
China (No. 19925522), the National Natural Science Foundation of
China (No. 90203001) and the Research Fund for the Doctoral
Program of Higher Education of China (Grant. No. 2000024832) and
ARC Large Grant A00103139 of the University of New South Wales
University, Australia.

\end{document}